\documentclass[conference,10pt]{IEEEtran}
\usepackage{epsfig,rotating,setspace,latexsym,amsmath,epsf,amssymb,amsfonts,bm,theorem,subfigure,epstopdf}
\usepackage{cite,authblk}
\usepackage{bbm}
\usepackage{algorithm}
\usepackage[noend]{algpseudocode}
\usepackage{color}
\usepackage{mathtools}

\algrenewcommand\algorithmicforall{\textbf{foreach}}
\algrenewcommand\algorithmicindent{.8em}

\IEEEoverridecommandlockouts
\allowdisplaybreaks

\begin{document}

\title{Timely Group Updating \thanks{This work was supported by NSF Grants CCF 17-13977 and ECCS 18-07348.}}

\author{Melih Bastopcu \qquad Sennur Ulukus\\
	\normalsize Department of Electrical and Computer Engineering\\
	\normalsize University of Maryland, College Park, MD 20742\\
	\normalsize \emph{bastopcu@umd.edu} \qquad \emph{ulukus@umd.edu}}

\maketitle

\begin{abstract}
We consider two closely related problems: anomaly detection in sensor networks and testing for infections in human populations. In both problems, we have $n$ nodes (sensors, humans), and each node exhibits an event of interest (anomaly, infection) with probability $p$. We want to keep track of the anomaly/infection status of all nodes at a central location. We develop a \emph{group updating} scheme, akin to group testing, which updates a central location about the status of each member of the population by appropriately grouping their individual status. Unlike group testing, which uses the expected number of tests as a metric, in group updating, we use the expected age of information at the central location as a metric. We determine the optimal group size to minimize the age of information. We show that, when $p$ is small, the proposed group updating policy yields smaller age compared to a sequential updating policy. 
\end{abstract}

\section{Introduction}
\begin{figure}[t]
\begin{center}
\subfigure[An anomaly detection system with multiple sensor nodes. Sensors in red indicate an anomaly and sensors in green indicate no anomaly.]
{\includegraphics[width=0.95\linewidth]{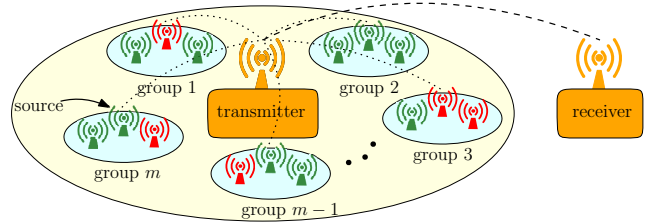}}\\ \vspace*{0.3cm}
\subfigure[An infection detection system in a human population. Persons in red are infected and persons in green are not infected.]
{\includegraphics[width=0.95\linewidth]{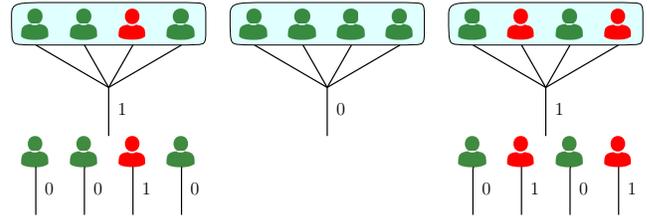}}
\end{center}
\vspace{-0.3cm}
\caption{System models considered in this paper.}
\label{fig:model}
\vspace{-0.4cm}
\end{figure}

We consider two different problems with similar system models: anomaly detection in sensor networks and testing for infections in human populations. In the anomaly detection problem, $n$ sensor nodes monitor a region and make measurements for an anomaly (e.g., fire, chemical spills, etc.) and report their measurements to a central location; see Fig.~\ref{fig:model}(a). Each sensor node detects an anomaly with probability $p$ independent of others. In the infection testing problem, there are $n$ individuals each of whom is infected with probability $p$ independent of others, and their infection status needs to be tallied at a central location; see Fig.~\ref{fig:model}(b). In both problems, we want to identify the anomaly/infection status of each node as timely as possible in order to take necessary actions as quickly as possible, e.g., control the fire or isolate/treat the infected persons. For a measure of timeliness, we use age of information, which keeps track of the time elapsed since the last time the status of a node is updated.

Inspired by the \emph{group testing} approach introduced in \cite{dorfman1943}, we develop a \emph{group updating} approach to maintain timely status updates at the central location. To that end, we divide $n$ nodes into groups of $k$ nodes each. In the case of anomaly detection, a local transmitter collects anomaly status of all nodes within the group. If there is no anomaly detected within the group, the local transmitter sends a single 0 to the central location. The central location, then, knows the status of all nodes within the group. On the other hand, if there is at least one anomaly detected within the group, the local transmitter sends a 1 to the central location. The central location, then, knows that there is at least one anomalous reading within the group. The local transmitter then sends the individual measurements of the sensors (0s and 1s) to the central location one-by-one. Similarly, in the case of testing humans for infection, we divide $n$ individuals into groups of $k$ each. Within each group, we mix the test samples of the individuals and perform a single test. If the test result is a 0, we know that no one within the group is infected. If the test result is a 1, then, we know that at least one person within the group is infected. In the latter case, we test each person within the group individually one-by-one. 

In the proposed group updating method, the group size $k$ plays an important role in the performance of the system, i.e., in the resulting age. If $k$ is too large, then the first update will likely result in a 1, and we will need to proceed to update the status of each node within the group one-by-one. This will increase the update duration, and hence, the age. On the other hand, if $k$ is too small, then this will result in too many groups, and therefore, too many updates within an update cycle. This will increase the age as well. Thus, there is an optimum group size $k$, which is not too small, not too large. In this paper, we determine that optimum size for given $n$ and $p$. 

Age of information has been used as a measure of timeliness in many communication and networking scenarios, such as, in web crawling, social networks, queueing networks, caching systems, energy harvesting systems, scheduling in networks, multi-hop multicast networks, lossless and lossy source coding, computation-intensive systems, vehicular, IoT, UAV systems, and so on \cite{Cho03, Ioannidis09, Costa16, Najm17, Sun17a, Soysal18, Yates17b, Bastopcu20d, Bastopcu20c, Farazi18, Wu18, Baknina18b, Leng19, Arafa19c, Arafa20a, Elmagid19, Kosta20, bastopcu_soft_updates_journal, Ceran18, Yates17a, Sun18a, Beytur18a, Kadota18a, Hsu18b, Maatouk20, Yavascan20, Buyukates19b, Buyukates18b, Mayekar20, MelihBatu4, Buyukates19c, Zou19a, Bastopcu19, Elmagid18}. With this paper, we are bringing age of information as a measure of timeliness to anomaly detection and testing for infections. Specifically relevant to our case is the setting of multi-source systems, where maximum age first (MAF) \cite{Sun18a}, maximum age difference (MAD) \cite{Beytur18a}, Whittle index \cite{Kadota18a, Hsu18b, Maatouk20}, slotted ALOHA with threshold \cite{Yavascan20}, hierarchical cooperation \cite{Buyukates19b} have been used to achieve good age performance. Different from most of these works, where only one source can be updated at a time, with the proposed group updating approach, we allow all sources in a group to be updated simultaneously with a single status update.

In this paper, we introduce a group updating approach where if all updates from the sources in the same group are 0, then the transmitter sends only a single status update representing the entire group, otherwise, the transmitter sends an update indicating that there is at least one 1 in the group, and proceeds to send all individual updates within the group one-by-one. For this updating method, for arbitrary $n$ and $p$, we first find an analytical expression for the average age, which depends on the group size $k$. For given $n$ and $p$, we find the optimal group size $k$ that minimizes the age. Next, we compare the performance of the proposed group updating policy with the performances of the traditional scheduling methods, and observe that the proposed group updating policy achieves a lower age than the existing schemes when $p$ is small. In addition, we compare the optimal group size $k$ in the group updating problem here and in the group testing problem in \cite{dorfman1943} and observe that they are different in general indicating the difference of the metrics used.           
\section{System Model} \label{sect:system_model}
We consider a system with $n$ sources/nodes. We divide the $n$ sources into groups of size $k$, where $m=\frac{n}{k}$ is the number of groups. Without loss of generality, we assume that $k$ divides $n$, and thus, $m$ is an integer. We denote the status of the $j$th source in the $i$th group in the $\ell$th update cycle by $X_{ij}(\ell)$, where $i=1,\ldots,m$, $j=1,\ldots,k$, and $\ell\geq 1$. $X_{ij}(\ell)$ is an independent and identically distributed (i.i.d.) binary random variable for all $i$, $j$ and $\ell$, with distribution,
\begin{align}
X_{ij}(\ell) = \begin{cases} 
1, & \text{with probability $p$}, \\
0, & \text{with probability $1-p$},
\end{cases}
\end{align}
where a status 1 indicates an anomaly/infection, and a status 0 indicates no anomaly/infection. 

Let $S_{ij}(\ell)$ denote the service time for the status update of the $j$th source in the $i$th group in the $\ell$th update cycle. This is the time it takes for the status of the node to go through the system and be tallied at the central location. Note that if the status of all nodes in the $i$th group is 0, then the service time for all nodes in this group is equal to 1, as in this case, for the anomaly detection problem, the local transmitter needs to send a single 0 to convey the status of all nodes, and in the infection testing problem, a single test will determine the infection status of all nodes in the group. On the other hand, if any one of the sources in the $i$th group generates 1 as a status update, the service time for the $j$th source in the $i$th group will be  equal to $j+1$, as in this case, an initial status update is sent representing the entire group, $j-1$ status updates are sent for the sources before source $j$, and a final update is sent for source $j$ itself. Thus, the service time for the $j$th node in group $i$ is a random variable with distribution,  
\begin{align}\label{service_eqn_1}
S_{ij}(\ell) = \begin{cases} 
1, & \text{with probability $(1-p)^k$}, \\
j+1, & \text{with probability $1-(1-p)^k$}.
\end{cases}
\end{align}

The service time of the entire $i$th group in the $\ell$th update cycle, denoted by $W_i(\ell)$, is equal to the service time of the last source in the $i$th group,  
\begin{align}\label{group_update_time}
W_i(\ell) = S_{ik}(\ell), \quad  i = 1,\dots,m.
\end{align}

As the central location wants to get timely updates from all sources, we track the age of each source at the central location separately. We denote the instantaneous age of source $j$ in group $i$ at time $t$ by $a_{ij}(t)$, with $a_{ij}(0)=0$. Age of each source at the central location increases linearly in time and drops to the age of the most recently received update once an update is received. The long term average age of node $j$ in group $i$ is given by,
\begin{align}\label{long_term}
   \Delta_{ij} = \lim_{T\to\infty} \frac{1}{T}\int_0^T a_{ij}(t)dt.
\end{align}

The overall average age of all sources $\Delta$ is equal to,
\begin{align}\label{total_age}
   \Delta = \frac{1}{n} \sum_{i=1}^{m}\sum_{j=1}^{k} \Delta_{ij}.
\end{align}
Our aim is to find the optimal group size $k^*$ that minimizes the average age of all sources $\Delta$, i.e., 
\begin{align}\label{problem1}
    k^* = \arg \min_{\{ k \}} \Delta.
\end{align}
In Section~\ref{Sec:Average_freshness}, we first find the average age, $\Delta$, in (\ref{total_age}).

\section{Average Age Analysis} \label{Sec:Average_freshness} 
\begin{figure}[t]
	\centering  \includegraphics[width=0.95\columnwidth]{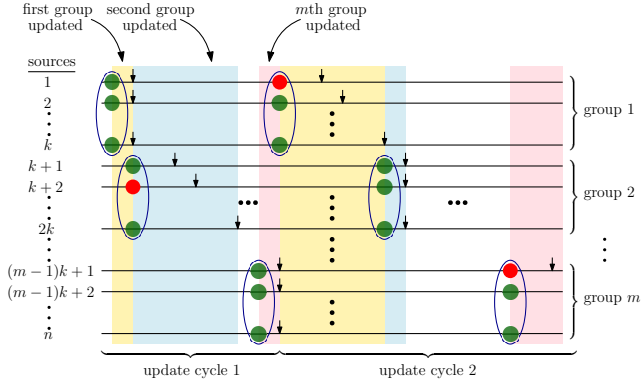}
	\caption{A sample update generation and update delivery timeline. Lines 1 through $n$ denote the nodes. Lines 1 through $k$ denote the nodes in group 1. Green and red balls represent the anomaly/no anomaly status of each node. In update cycle 1, the yellow strip shows the time where the status of all nodes in group 1 is updated, the blue strip shows the time where the status of all nodes in group 2 is updated, and the pink strip shows the time where the status of all nodes in group $m$ is updated. The process repeats itself in update cycle 2. Delivery times are marked by the downward arrows.}
	\label{fig:uptEvol}
	\vspace{-0.4cm}
\end{figure}

With the group updating policy, the transmitter starts with sending updates from the sources in the first group. If all the updates from the first group are 0 (as shown with green balls in the first $k$ lines in Fig.~\ref{fig:uptEvol}), then the transmitter sends a single 0 to update all the sources in the first group (that is why the delivery times of updates for all sources in the first group marked with arrows in Fig.~\ref{fig:uptEvol} are equal to 1). After sending updates from the first group, the transmitter proceeds to send updates from the second group. If any one of the updates from the second group is equal to 1 (denoted by a red ball in the lines between lines $k+1$ and $2k$ in Fig.~\ref{fig:uptEvol}), then the transmitter first sends a 1 as a status update representing the entire group, and then sends individual updates from each source one-by-one. As shown in Fig.~\ref{fig:uptEvol}, the receiver gets the first update from the second group after 2 units of time. After sending updates from the second group, the transmitter proceeds to send updates from the third group, and so on, up until the $m$th (last) group. We call this entire time in which the status of all $n$ sources are updated as update cycle 1 in Fig.~\ref{fig:uptEvol}. Once update cycle 1 ends, update cycle 2 starts all over again with all sources taking a new i.i.d.~realization. In Fig.~\ref{fig:uptEvol}, in update cycle 1, the yellow vertical strip shows the time in which the status of all nodes in group 1 is updated, the blue strip shows the time in which the status of all nodes in group 2 is updated, so on so forth, and finally, the pink strip shows the time in which the status of all nodes in group $m$ is updated.         

\begin{figure}[t]
	\centering  \includegraphics[width=0.95\columnwidth]{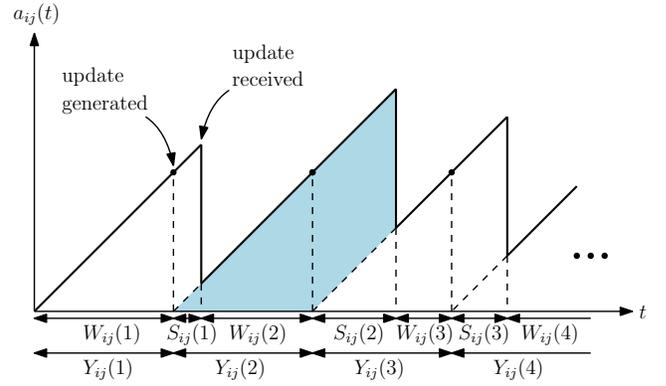}
	\caption{A sample age evolution $a_{ij}(t)$ at the central location.}
	\label{fig:ageEvol}
	\vspace{-0.4cm}
\end{figure}

Fig.~\ref{fig:ageEvol} shows a sample age evolution curve for the $j$th source in the $i$th group at the central location, i.e., $a_{ij}(t)$. Here, $S_{ij}(\ell)$ defined in (\ref{service_eqn_1}) denotes the service time of the $j$th source in the $i$th group in the $\ell$th update cycle. In addition, $W_{ij}(\ell)$ denotes the total waiting time until the $\ell$th update is generated after the service completion of the $(\ell-1)$th update for the same source. Thus, $W_{ij}(\ell)$ is given by,     
\begin{align}\label{waiting_eqn_1}
W_{ij}(\ell) = \bar{W}_{ij}(\ell-1)+\sum_{r=i+1}^{m}W_r(\ell-1) +\sum_{r=1}^{i-1}W_r(\ell),
\end{align}
where $W_r(\ell)$ is given in (\ref{group_update_time}), and $\bar{W}_{ij}(\ell-1)$ denotes the remaining service time of the $i$th group in the ($\ell-1$)th update cycle which is given by $\bar{W}_{ij}(\ell-1) = W_i(\ell-1)- S_{ij}(\ell-1)$. 

We denote the length of the $\ell$th update cycle for the $j$th source in the $i$th group as $Y_{ij}(\ell) = S_{ij}(\ell-1)+W_{ij}(\ell)$ with $S_{ij}(0) = 0$ for convention. One can show that the long term average age $\Delta_{ij}$ given in (\ref{long_term}) as in \cite{Najm17} is, 
\begin{align}
	\Delta_{ij} &= \lim_{N\to\infty} \frac{\frac{1}{N}\left(\frac{1}{2}\sum_{\ell=1}^{N+1} Y_{ij}(\ell)^2+\sum_{\ell=1}^{N} Y_{ij}(\ell)S_{ij}(\ell) \right)}{\frac{1}{N}\sum_{\ell=1}^{N}Y_{ij}(\ell)},\label{avg_age1}
\end{align}
where $N$ denotes the number of update cycles. We note that (\ref{avg_age1}) can be written equivalently as,
\begin{align}
    \Delta_{ij} =  \frac{\mathbb{E}[Y_{ij}^2]}{2\mathbb{E}[Y_{ij}]}+\mathbb{E}[S_{ij}]. \label{avg_age2}
\end{align}     

We note that the length of an update cycle $Y_{ij}$ is equal to the service completion time of all the groups, i.e.,
\begin{align} \label{Yij-in-Wr}
Y_{ij} = S_{ij}+W_{ij}=\sum_{r=1}^{m} W_r.
\end{align}
Therefore, the variable $Y_{ij}$ does not depend on $i$ or $j$. We thus denote $Y_{ij}$ with a single random variable $Y$, i.e., $Y=Y_{ij}$. On the other hand, from (\ref{service_eqn_1}), $S_{ij}$ depends on $j$, and we denote it by $S_j$. Then, the overall average age $\Delta$ in (\ref{total_age}) is equal to, 
\begin{align}\label{total_age2}
    \Delta = \frac{\mathbb{E}[Y^2]}{2\mathbb{E}[Y]}+\mathbb{E}[S],
\end{align}
where $\mathbb{E}[S] =\frac{1}{n}\sum_{i=1}^{m}\sum_{j=1}^{k} \mathbb{E}[S_{ij}]=\frac{1}{k}\sum_{j=1}^{k} \mathbb{E}[S_{j}]$.

Now, using (\ref{service_eqn_1})-(\ref{group_update_time}), we have $\mathbb{E}[W_{r}] = 1+k\left( 1-(1-p)^k\right)$ for all $r$. Thus, from (\ref{Yij-in-Wr}),
\begin{align}\label{exp_y_ij}
    \mathbb{E}[Y] =& \frac{n}{k}+n\left( 1-(1-p)^k\right).
\end{align}
In addition, 
\begin{align}\label{exp_y_ij2}
    \mathbb{E}[Y^2] 
    =& n(n-k)(1-p)^{2k}+\frac{n^2(k+1)^2}{k^2}\nonumber\\
    &- n\left(2n\left(1+\frac{1}{k}\right)-k\right)(1-p)^k.
\end{align}    

Further, from (\ref{service_eqn_1}), we have, 
\begin{align}\label{exp_s_ij}
    \mathbb{E}[S_{ij}] = \mathbb{E}[S_{j}] = 1+j(1-(1-p)^k),
\end{align}
and thus, we have,
\begin{align}\label{exp_s}
    \mathbb{E}[S] = 1+\frac{k+1}{2}(1-(1-p)^k).
\end{align}

Hence, the overall average age  $\Delta$ in (\ref{total_age2}) is  
\begin{align}\label{total_age3}
    \Delta =& \frac{k^2(n-k)(1-p)^{2k}+n(k+1)^2}{2k+2k^2\left( 1-(1-p)^k\right)}\nonumber\\
    &-\frac{\left(2n\left(k+1\right)-k^2\right)(1-p)^k}{2+2k\left( 1-(1-p)^k\right)}\nonumber\\
    &+1+\frac{k+1}{2}(1-(1-p)^k).
\end{align}

The overall average age in (\ref{total_age3}) depends on $n$, $p$ and $k$. We find the optimal $k$ that minimizes $\Delta$ numerically in Section~\ref{sect:num_res}. 

\section{Group Updating versus Group Testing} \label{sect:comparison}
While the group testing and group updating policies are operationally similar, parameter selection, mainly selection of the group size in both problems, is different. In particular, in group testing, group size $k$ is chosen to minimize the expected number of tests. In our terminology, expected number of tests corresponds to the expected length of an update cycle, i.e., $\mathbb{E}[Y]$, as the transmitter sends one status update at a time. Thus, group testing chooses the group size $k_{gt}^*$ by solving, 
\begin{align}\label{problem1_group}
    k_{gt}^* = \arg \min_{\{ k\in \mathbb{Z}^+ \}} \mathbb{E}[Y],
\end{align}
where $\mathbb{E}[Y]$ is given in (\ref{exp_y_ij}). In order for group testing to be more efficient than sequential updating of sources one-by-one, which uses $n$ tests in an update cycle, we need $\mathbb{E}[Y]\leq n$, which implies $p\leq p_{gt}$, where
\begin{align}
    p_{gt} = 1-\left(\frac{1}{k}\right)^{\frac{1}{k}}.
\end{align}
We note that $p_{gt}$ attains its maximum value $0.3066$ when $k=3$. Thus, when $p >0.3066$, group testing becomes inefficient compared to sequential updating of sources one-by-one. 

Next, we find $k_{gt}^*$ in (\ref{problem1_group}) analytically. For that, we first relax the integer constraint on $k$. Then, by equating the derivative of $ \mathbb{E}[Y]$ in (\ref{exp_y_ij}) with respect to $k$ to zero, we obtain,
\begin{align}\label{derivative}
    \frac{\partial \mathbb{E}[Y]}{\partial k } = -\frac{n}{k^2}-n(1-p)^k\log(1-p) =0,
\end{align}
which gives, 
\begin{align}\label{derivative2}
    \frac{k}{2}\log(1-p)e^{\frac{k}{2}\log(1-p)} =-\frac{1}{2}\sqrt{-\log(1-p)}.
\end{align}
Note that (\ref{derivative2}) is in the form of $ xe^x=y$, whose solutions for $x$ are $x_1 = W_{0}(y)$ and $x_2 = W_{-1}(y)$ when $-\frac{1}{e}\leq y< 0$. Here, $W_0(\cdot)$ and $W_{-1}(\cdot)$ denote the principle and $-1$st branches of the Lambert $W$ function, respectively \cite{lambert}. Thus, when $0<p\leq 1-e^{-\frac{4}{e^2}} = 0.418$, we have two solutions for (\ref{derivative2}) which are given by,
\begin{align}
    \alpha_1 &= \frac{2}{\log(1-p)} W_{0}\left( -\frac{1}{2}\sqrt{-\log(1-p)}\right),\label{root1}\\
    \alpha_2 &= \frac{2}{\log(1-p)} W_{-1}\left( -\frac{1}{2}\sqrt{-\log(1-p)}\right).\label{root2}
\end{align}
When $p>0.418$, one can show that $\frac{\partial \mathbb{E}[Y]}{\partial k }<0$, and thus, the optimal $k$ is equal to $n$. However, as the group testing method becomes inefficient when $p>0.3066$, we only need to consider the case when $0<p\leq 0.418$, and thus, $\alpha_1$ in (\ref{root1}) and $\alpha_2$ in (\ref{root2}) always exist.  

Thus, in order to find the optimal $k$, we need to check $k=\alpha_r^{u}$ where $\alpha_r^{u}= \min \{k | k\geq \alpha_r, k | n \} $ for $r=1,2$; $k=\alpha_r^{\ell}$ where $\alpha_r^{\ell}= \max \{k | k\leq \alpha_r, k | n\}$ for $r=1,2$; $k=1$; and $k=n$. In other words, the optimal $k$ is given by, 
\begin{align}\label{problem1_group_2}
    k_{gt}^* = \arg \min_{\{ k\in \mathbb{K} \}} \mathbb{E}[Y],
\end{align}
where $\mathbb{K}=\{1, \alpha_1^{\ell},\alpha_2^{\ell}, \alpha_1^{u},\alpha_2^{u}, n \}$. 

We perform a similar analysis for the group updating problem. Group updating chooses the group size $k_{gu}^*$ by solving, 
\begin{align}\label{problem1_group_update}
    k_{gu}^* = \arg \min_{\{ k\in \mathbb{Z}^+ \}} \Delta,
\end{align}
where $\Delta$ is given in (\ref{total_age3}). In order for group updating to be more efficient than sequential updating, $\Delta$ in (\ref{total_age3}) needs to be smaller than $\Delta_{\text{round-robin}}$. For the round-robin (sequential) scheduling method, $\mathbb{E}[Y] = n$, $\mathbb{E}[Y^2] = n^2$, $\mathbb{E}[S] = 1$, and the overall average age from (\ref{total_age2}) is,
\begin{align}\label{age_round}
 \Delta_{\text{round-robin}}=\frac{n}{2}+1.  
\end{align}
The condition $\Delta\leq \Delta_{\text{round-robin}}$ gives an upper bound for the probability $p$, which we denote by $p_{gu}$. In other words, when $p>p_{gu}$, group updating becomes inefficient compared to sequential updating. Further, by relaxing the integer constraint on $k$ and equating the derivative of $\Delta$ in (\ref{total_age3}) with respect to $k$ to $0$, we can find the critical points where the age is minimized, and find $k_{gu}^*$ analytically. Since $\Delta$ in (\ref{total_age3}) is an involved function of $k$, in this work, we do not pursue analytical results on $p_{gu}$ and $k_{gu}^*$. Instead, we find $k_{gu}^*$ for given of $p$ and $n$, and examine $p_{gu}$, numerically, in the next section.  

\section{Numerical Results} \label{sect:num_res}
In this section, we provide four numerical results to illustrate the performance of the proposed group updating method, and also to show its difference from the group testing method. In all the numerical results, we only consider $k$ values that divide $n$. For example, if $n = 6$, we consider $k = 1,2,3,6$.  

In the first numerical example, we compare the performance of the proposed group updating method with the performances of the existing updating policies of MAF and MAD. Since after receiving each update, the age at the receiver goes down to 1, MAF and MAD scheduling policies become identical. In addition, as the ages of all sources start from zero, MAF and MAD policies become the same as the round-robin scheduling method. The average age for the round-robin scheme is given in (\ref{age_round}). We note that $\Delta_{\text{round-robin}}$ increases linearly with $n$ and does not depend on the probability $p$. 

\begin{figure}[t]
	\centerline{\includegraphics[width=0.9\columnwidth]{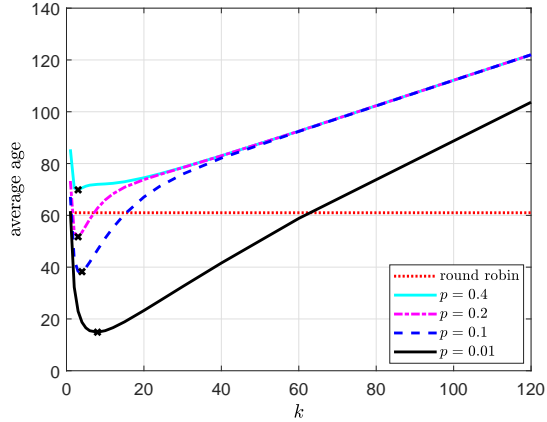}}
	\vspace{-0.3cm}
	\caption{Average age versus group size with the proposed group updating method and the round robin method when $p = 0.01,0.1,0.2,0.4$.}
	\label{Fig:sim1}
	\vspace{-0.4cm}
\end{figure}

In the first numerical example, we take $n=120$ and plot in Fig.~\ref{Fig:sim1} the average age $\Delta$ in (\ref{total_age3}) with respect to $k$ when $p=0.01,0.1,0.2,0.4$, together with $\Delta_{\text{round-robin}}$ in (\ref{age_round}). We observe in Fig.~\ref{Fig:sim1} that, for all values of $p$, the average age first decreases with $k$ and then increases with $k$, as initially, increasing $k$ decreases the number of groups, making group updating more efficient, but after a while, further increasing $k$ decreases the likelihood of having all zero updates in a group, requiring many follow-up individual updates. Thus, there is a trade-off between these two opposing factors, and there is an optimum group size to minimize the average age. As marked with a cross in Fig.~\ref{Fig:sim1}, when $p=0.01$ the optimal group size is $k_{gu}^*=8$; when $p=0.1$ it is $k_{gu}^*=4$; when $p=0.2$ it is $k_{gu}^*=3$; and when $p=0.4$ it is $k_{gu}^*=3$. We also observe that the group updating method becomes inefficient with increased $p$ as it becomes more likely for the transmitter to send individual updates. When $p$ is large enough, e.g., when $p = 0.4$, we observe in Fig.~\ref{Fig:sim1} that group updating becomes inefficient and does not improve the average age compared to the round-robin scheduling method.

\begin{figure}[t]
	\centerline{\includegraphics[width=0.9\columnwidth]{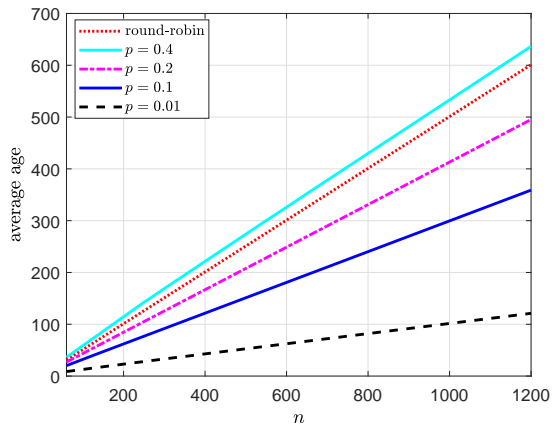}}
	\vspace{-0.3cm}
	\caption{Average age versus population size with the proposed group updating method and the round robin method when $p=0.01,0.1,0.2,0.4$.}
	\label{Fig:sim2}
	\vspace{-0.4cm}
\end{figure}

In the second numerical example, we again take $p=0.01,0.1,0.2,0.4$, and plot in Fig.~\ref{Fig:sim2} the average age with respect to $n$, the population size, for $n$ from $60$ to $1200$. For each value of $p$ and $n$, we first find the optimal $k_{gu}^*$ that achieves the minimum age, then plot that minimum age with respect to $n$. We observe in Fig.~\ref{Fig:sim2} that the average age increases linearly with the proposed group updating method as with the round-robin scheduling method. Similar to the first numerical example, the average age increases with $p$ as group updating becomes less efficient with larger $p$.

\begin{figure}[t]
	\begin{center}
	\subfigure[]{%
	\includegraphics[width=0.9\linewidth]{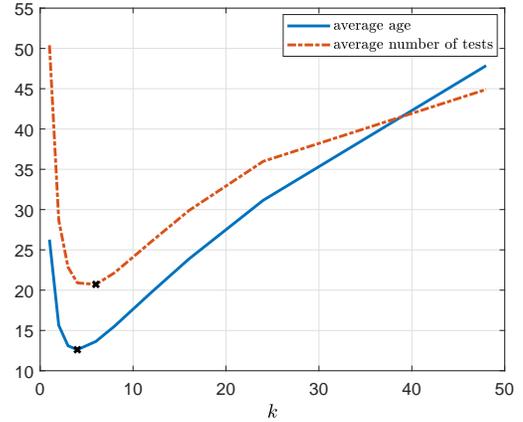}}\\\vspace{-0.3cm}
	\subfigure[]{%
	\includegraphics[width=0.9\linewidth]{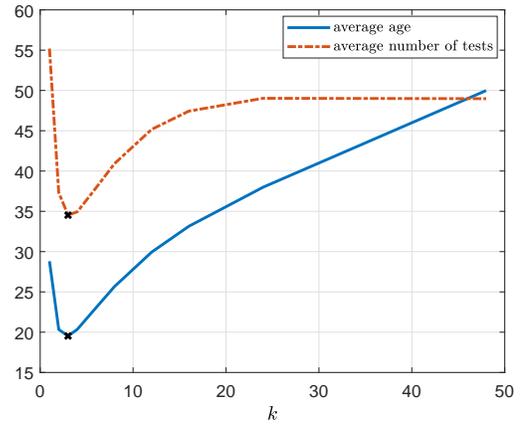}}
	\end{center}
	\vspace{-0.3cm}
	\caption{Average age for the group updating method and average number of updates for the group testing method with respect to $k$ for $n=48$ when (a) $p=0.05$ and (b) $p= 0.15$.}
	\label{Fig:sim3}
	\vspace{-0.4cm}
\end{figure}

In the third numerical example, we examine the differences between the group updating problem and the group testing problem. For this numerical example, we take $n = 48$, $p = 0.05, 0.15$, and determine the optimal $k$ values that minimize the average age and also the average number of updates. When $p$ is small, e.g., when $p = 0.05$, we observe in Fig.~\ref{Fig:sim3}(a) that the optimal group size that minimizes the average age is $k_{gu}^*=4$, whereas the optimal group size that minimizes the average number of updates is  $k^*_{gt}=6$. This verifies that the group updating problem is different than the group testing problem. However, when $p$ is relatively large, e.g., when $p=0.15$, we observe in Fig.~\ref{Fig:sim3}(b) that the optimal group sizes in both problems are equal $k_{gu}^*=k_{gt}^*=3$. In other words, when $p$ gets larger, the optimal $k$ values for the group updating and group testing problems get closer to each other.

\begin{figure}[t]
	\centerline{\includegraphics[width=0.9\columnwidth]{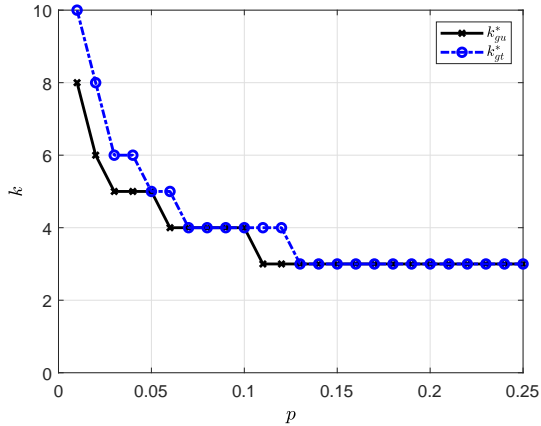}}
	\vspace{-0.3cm}
	\caption{Optimum group sizes $k_{gu}^*$ in the group updating problem and $k^*_{gt}$ in the group testing problem for $n=120$, for $p$ from $0.01$ to $0.25$.}
	\label{Fig:sim4}
	\vspace{-0.4cm}
\end{figure}

In the fourth numerical example, we examine $k_{gu}^*$ and $k_{gt}^*$ as a function of $p$. We take $n=120$ and vary $p$ between $0.01$ and $0.25$. We observe in Fig.~\ref{Fig:sim4} that both $k_{gu}^*$ and $k^*_{gt}$ decrease with probability $p$. With higher $p$, the sources in a group begin to generate more $1$s as status updates, which results in sending more individual updates from the sources. Thus, decreasing the group size $k^*$ in both of the problems helps counter the effects of increased $p$. Similar to the previous example, we observe in Fig.~\ref{Fig:sim4} that $k_{gu}^*$ and $k^*_{gt}$ are different when $p$ is small, and become the same when $p\geq 0.13$ for this choice of $n$.        

\section{Conclusion}
We considered the problem of timely group updating, where similar to group testing, the sources are divided into groups; if all updates within a group are negative, a single group update suffices; if at least one update is positive, this triggers a sequence of individual updates. For this updating scheme, we derived an analytical expression for the average age as a function of the group size $k$, the number of sources $n$, and the probability $p$. For given $n$ and $p$, we found the optimal group size $k$ that minimizes the age. We showed that when $p$ is small, group updating performs better than sequential updating. We also showed that the optimal group sizes for group updating and group testing are different. This is because, while group testing aims to minimize the first moment of the length of an update cycle, group updating aims to minimize the age which depends on both the first and second moments of the length of an update cycle. An analogous observation was made in timely source coding versus traditional source coding, where the former depends on the first and second moments of the codeword length, while the latter depends only the first moment \cite{Mayekar20, MelihBatu4}.

\bibliographystyle{unsrt}
\bibliography{IEEEabrv,lib_v1_melih}
\end{document}